\documentstyle[12pt]{article}
\begin{document}
\renewcommand{\theequation}{\arabic{section}.\arabic{equation}}
\def\sq{\hbox{\vrule\vbox{\hrule\phantom{o}\hrule}\vrule}}
\def\wh{\widehat}
\def\wt{\widetilde}
\def\ovr{\overrightarrow}
\def\ol{\overleftarrow}
\def\om{\omega}
\def\ov{\overline}
\def\noi{\noindent}
\def\ep{\varepsilon}
\font\cincorm=cmr5
\font\ninerm=cmr9
\font\docerm=cmr12
\font\bigbf=cmbx10 scaled 1440
\mathsurround=1pt
\begin{titlepage}

Preprint \hfill   \hbox{\bf SB/F/99-263}
\hrule
\vskip 1.5cm

\centerline{\bf NON ABELIAN TQFT AND SCATTERING OF}
\centerline{\bf SELF DUAL FIELD CONFIGURATIONS}
\vskip 1cm

\begin{description}
\item[] \centerline {R. Gianvittorio, A. Restuccia and J.F. S\'anchez}
\item[] \centerline {\it  Universidad Sim\'on
Bol\'{\i}var,Departamento de F\'{\i}sica,}
\item[] \centerline {\it Apartado Postal 89000,  Caracas 1080-A, Venezuela.}
\item[] \centerline{\it \ \ \ e-mail: ritagian@usb.ve, arestu@usb.ve, jsanchez@fis.usb.ve}
\end{description}
\vskip 1cm

\centerline{\bf Abstract}

\vskip .5cm

A non-abelian topological quantum field theory describing the scattering of self-dual field configurations over topologically non-trivial Riemann surfaces, arising from the reduction of 4-dim self-dual Yang-Mills fields, is introduced. It is shown that the phase space of the theory can be exactly quantized in terms of the space of holomorphic structures over stable vector bundles of degree zero over Riemann surfaces. The Dirac monopoles are particular static solutions of the field equations. It´s relation to topological gravity is discussed.

\vskip 2cm
\hrule
\bigskip
\centerline{\bf UNIVERSIDAD SIMON BOLIVAR}
\vfill
\end{titlepage}

\setcounter{page}{2}
\section{INTRODUCTION}
\setcounter{equation}{0}

We introduce  a  non-abelian topological field theory (TFT), which
extends the BF class of TFT  \cite{BT}-\cite{BBRT}.  The theory we
propose is constructed over a  three dimensional space-time with boundaries
being Riemann surfaces, which will be interpreted as incoming and
outgoing ``branes''.  Although we discuss only the case of a 3-dim
base manifold, the topological action we present may be extended to other
dimensions as occurs with the BF theories.  One motivation of
our proposal is to describe a TFT whose path integral, with boundary
conditions determined by functionals $\psi_{in}$ and $\psi_{out}$ on
the space of self-dual Yang-Mills fields dimensionally reduced to
2-dimensions, at
the incoming and outgoing branes $M_{in}$ and $M_{out}$, may be interpreted as a
scattering amplitude \cite{A}-\cite{W} of self-dual fields.
Alternatively the path integral may  be
used to define a map from functionals of the fields $\varphi$ at $M_{in}$
to the fields at $M_{out}$ by considering
\begin{equation}
\label{path}
\psi_{out}(\varphi_{out})= \int_{\varphi |_{Mout}=\varphi_{out}} {\cal D}
\varphi \;
exp(-S_{eff}) \; \psi_{in}(\varphi_{in}) \; ,
\end{equation}
where the integral is performed over all fields whose restriction to
$M_{out}$ is equal to $\varphi_{out}$. In our proposal the fields
$\varphi_{in}$ and $\varphi_{out}$ will correspond to solutions of the
Hitchin ``self-dual'' equations over Riemann surfaces \cite{H}
arising from the dimensional reduction of 4-dim self-dual
Yang-Mills connections. Since
the irreducible solutions of these equations are in one to one correspondence to the holomorphic structures over stable vector bundles on Riemann surfaces, the path
integral (\ref{path}) provides a map between functionals of the
holomorphic structures at the boundaries $M_{in}$ and $M_{out}$.

Since one sector of the irreducible solutions of Hitchin equations provides a model of Teichm\"{u}ller space as a $(3g-3)$ dimensional complex space, our proposal should be related to topological gravity in 3-dim \cite{Ge}. In fact, it is known that $(2+1)$ dimensional gravity can be exactly quantized on a space-time with topology $R\times\Sigma$ where $\Sigma$ is a genus $g$ Riemann surface. Its phase space has the structure of a cotangent bundle. More precisely it is the cotangent bundle
\[
T^{*}({\cal A/G})
\]
where ${\cal A}$ is the space of flat $SO(2,1)$ connections over $\Sigma$, and ${\cal G}$ is the gauge group of topological gravity in 3-dim. The space ${\cal A/G}$ is isomorphic to Teichm\"{u}ller space, hence the physical Hilbert space of states is the space of square integrable functions over Teichm\"{u}ller space (or the moduli space of Riemann surfaces when diffeomorphisms not isotopic to the identity are included). We will show that the physical phase space of the TQFT we propose can be exactly quantized, we will then be able to determine the relation between the Hilbert space of physical states in $(2+1)$ gravity and our proposal,  constructed from the reduction of 4-dim self-dual Yang-Mills fields. Previously to the presentation of the TQFT we discuss a geometric interpretation of BF TFT and its natural generalization in terms of Hitchin self-dual equations. We will show later on that these equations arise as first class constraints on the TFT we present.

\section{Geometrical Interpretation of BF theories}
\setcounter{equation}{0}

If $V$ is a $C^{\infty}$ vector bundle over a Riemann surface $\Sigma$, a
holomorphic structure on $V$ is a differential operator
\begin{equation}
d''_{B}: \Omega^{0}(\Sigma ,V) \longrightarrow \Omega^{0, 1}(\Sigma ,V)
\end{equation}
such that
\begin{equation}
d''_{B}(f s) = d''f \otimes s + f \;
d''_{B}s \ ,
\end{equation}
for any $C^{\infty}$ function $f$ and section $s \; \epsilon \; \Omega^{0}(\Sigma ,V) $.
We denote by $\Omega^{p}(\Sigma ,V) $ the p-forms on $\Sigma$ with values in $V$.
This definition is equivalent to the more usual one given in terms of
local trivializations which are related by holomorphic transitions
functions \cite{H}.

Two holomorphic structures are said to be equivalent if there is a
gauge transformation such that
\begin{equation}
g^{-1} d''_{B} g = d''_{B_{2}} .
\end{equation}
Locally, we can write
\begin{equation}
d''_{B}=d'' + B d\ov{z}
\end{equation}
where $ d''f= \frac{ \partial f}{\partial \ov{z}}d\ov{z}$ and $B=B(z, \ov{z} )$.

When the vector bundle $V$ has hermitian structure we can
associate to $d''_{B}$ a unitary connection $A$. Locally it may be
expressed as
\begin{equation}
A=B d\ov{z} - B^{*} dz .
\end{equation}
From this point of view, the space of holomorphic structures on $V$
may be analysed in terms of the space of unitary connections. This is
the starting point in the analysis of Hitchin in \cite{H}.

A first result relating both spaces over a trivial line bundle $\Sigma
\times C$ is that every holomorphic structure  on $\Sigma
\times C$  is equivalent to the holomorphic structure of a flat unitary
connection. This connection is unique modulo unitary gauge
transformations.

This result was generalized by Narashmhan and Seshradi to vector
bundles of higher rank: every holomorphic structure on a vector
bundle over $\Sigma$ of degree zero which is stable, is equivalent to the
holomorphic structure of a flat irreducible unitary connection. The
connection is unique modulo unitary gauge transformations.

This theorem stablishes then the relation between the space of stable
holomorphic structures over $V$ and unitary BF ``topological'' theories.
In fact the latest describe the geometry of flat connections over
$(V,\Sigma )$. The one to one correspondence is then obtained provided the
BF theory describes flat irreducible unitary connections.

There is a generalization of the notion of geometrical stability for
rank 2 holomorphic vector bundles of degree zero. It was introduced by
Hitchin in \cite{H}, and refers to the stability of pairs $(V, \Phi)$
where $\Phi$ is a holomorphic section of $ End V \otimes K$ ($K$ is the canonical bundle). The
generalization of the theorem of Narashmhan and Seshradi was
formulated in \cite{H} and stablishes that the holomorphic structure
of a rank 2 holomorphic vector bundle $V$ of degree zero over a Riemann
surface $\Sigma$ of genus $g > 1$, such that $(V, \Phi)$ is stable in the
sense of Hitchin, is equivalent to an irreducible solution of the
``self-duality'' equations:

\begin{eqnarray}
\label{hit1}
F_{A} + [ \Phi, \Phi^{*}] =0 , \\
\label{hit2}
d''_{A} \Phi=0 .
\end{eqnarray}

The solution is unique modulo unitary gauge transformations. These
equations were called ``self-dual'' because they arise also from the
4-dimensional self dual condition
\begin{equation}
F_{A}=^{*}F_{A} ,
\end{equation}
by dimensional reduction.

They are of course globally defined over $\Sigma$ and are conformal
invariant.

The topological field theory describing the Hitchin field equations
over Riemann surfaces was
obtained in \cite{MR}. It gives then a generalization of the BF
theory over Riemann surfaces.

One of the interesting properties of the ``self-dual'' equations over
Riemann surfaces is that one sector of the full moduli space of
solutions provide a model of Teichm\"uller space as a full $(3g-3)$
dimensional complex space.

The topological action we will introduce in the next section
formulated over a 3-dimensional space time will allow to map
functionals over the space of solutions of the Hitchin equations at
an initial time, describing an holomorphic structure over the Riemann
surface of a given topology to solutions at a final time describing
some other holomorphic structure, in terms of an $SU(2)$ connection
$A$ and a $SU(2)$ valued 1-form $\Phi$.

\section{The gauge invariant action}
\setcounter{equation}{0}

The topological $SU(2)$ invariant action we propose is

\begin{equation}
 S = -2\int_{M} d^3x \; \epsilon^{\sigma\mu\nu} \; {\rm
Tr}\left [B_\sigma(F_{\mu\nu} + [\phi_\nu,\phi_\mu])
+ \eta_\sigma D_{[\mu}\phi_{\nu]}\right ] \; ,
\end{equation}
where $M$ a 3 dimensional manifold $\Sigma \times K$ where $\Sigma$ is a Riemann Surface of genus $g$ and $\mu , \nu , \sigma = 0,1,2 $.

\noi  The field  F is the Lie algebra valued 2-form curvature
corresponding to
 the 1-form gauge connection A; B,$ \phi$ and $\eta$ are independents
Lie
algebra valued 1-forms. The covariant derivative is defined as $D_\mu
w = \partial_\mu w + [A_\mu , w]$.

After  expanding the fields in terms of the $SU(2)$ generators
$T_{a} \; (a=1,2,3)$¥,  the action may be rewritten  as
\begin{equation}
\label{action JF}
S = \int_{M} d^3 x \; \epsilon^{\sigma\mu\nu}\left [{B_\sigma}\!^a
({F_{\mu\nu}}\!^a + \epsilon^{abc}{\phi_\nu}\!^b {\phi_\mu}\!^c) +
{\eta_\sigma}\!^a D_{[\mu}{\phi_{\nu]}}^a \right ] \; .
\end{equation}

\noi Taking variations with respect to the gauge field $A^a_\mu$ we
obtain
the field equation
\begin{equation}
\label{feq}
 \epsilon^{\sigma\mu\nu}(\partial_\mu {B_\sigma}\!^a -
\epsilon^{abc}({B_\sigma}\!^b {A_\mu}\!^c + {\eta_\sigma}\!^b
{\phi_\mu}\!^c)) = 0 \; ,
\end{equation}
 variations with respect to the field $\phi^a_\mu$ yields
\begin{equation}
\label{feqq}
 \epsilon^{\sigma\mu\nu} (\partial_\mu {\eta_\sigma}\!^a +
\epsilon^{abc} ({A_\mu}\!^b {\eta_\sigma}\!^c + {B_\sigma}\!^b
{\phi_\mu}\!^c)) = 0 \; ,
\end{equation}
variations with respect to the field  $B^a_\mu$ gives
\begin{equation}
\label{feq1}
{F_{\mu\nu}}\!^a + \epsilon^{abc} {\phi_\nu}\!^b {\phi_\mu}\!^c = 0  ,
\end{equation}
finally, variations with respect to the field
 $\eta^a_\mu$ leads to
\begin{equation}
\label{feq2}
D_{[\mu} {\phi_{\nu]}}\!^a = 0  .
\end{equation}

It can be shown after some calculations that (\ref{feq1}) and
(\ref{feq2}) are
the integrability conditions for equations (\ref{feq}) and (\ref{feqq}).
Consequently in order to solve the field equations, we may first look
for solutions of (\ref{feq1}) and (\ref{feq2}) which then  ensures
the existence of
solutions for $B$ and $\eta$.

The action (\ref{action JF}) has several gauge symmetries which we will discuss
shortly. They are given by

\begin{eqnarray}
\delta {A_\mu}\!^a & = & - D_\mu {\varepsilon_1}\!^a  +
\epsilon^{abc} {\phi_\mu}\!^b \, {\varepsilon_2}\!^c  , \nonumber \\
\delta {\phi_\mu}\!^a & = & - D_\mu {\varepsilon_2}\!^a -
\epsilon^{abc} {\phi_\mu}\!^b \, {\varepsilon_1}\!^c  , \nonumber \\
\delta {\eta_\mu}\!^a & = & - D_\mu {\varepsilon_3}\!^a +
\epsilon^{abc} {\phi_\mu}\!^b \, {\varepsilon_4}\!^c - \epsilon^{abc}
{\eta_\mu}\!^b \, {\varepsilon_1}\!^c + \epsilon^{abc} {B_\mu}\!^b \,
{\varepsilon_2}\!^c  ,\nonumber \\
\label{symm}
\delta {B_\mu}\!^a    & = & - D_\mu {\varepsilon_4}\!^a -
\epsilon^{abc} {\phi_\mu}\!^b \, {\varepsilon_3}\!^c - \epsilon^{abc}
{\eta_\mu}\!^b \, {\varepsilon_2}\!^c - \epsilon^{abc} {B_\mu}\!^b \,
{\varepsilon_1}\!^c  .
\end{eqnarray}

The action  (\ref{action JF}) formulated in a canonical form,
may be expressed as
\begin {eqnarray}
\label{canon}
S =  \int_{M}  d^3x  \left [ \pi^{ia} {{\dot{A}}_i}\!^a +
 P^{ia} {{\dot{\phi}}_i}\!^a + {A_0}\!^a (D_i \pi^{ia} +
\epsilon^{abc} {\phi_i}\!^b P^{ic}) \right . \nonumber \\
+{\phi_0}\!^a(D_i P^{ia} - \epsilon^{abc} {\phi_i}\!^b \pi^{ic})
 +{\eta_0}\!^a D_{[i} {\phi_{j]}}^a \epsilon^{oij} \nonumber \\
 +\left . {B_0}\!^a ({F_{ij}}\!^a + \epsilon^{abc} {\phi_j}\!^b
{\phi_i}\!^c)\epsilon^{oij}\right ]  .
\end{eqnarray}
From (\ref{canon}) we recognize a constrained theory with vanishing canonical
Hamiltonian,
where the canonical conjugate momenta associated to $A_i^a$ is given by
\begin {equation}
\pi^{ia}=2 \epsilon^{0ij} B_j^a  ,
\end {equation}
while the one associated to $\phi_i^a$ is
\begin {equation}
P^{ia}=2 \epsilon^{0ij}\eta_j^a  .
\end {equation}
The fields $A_0^a, \phi_0^a, \eta_0^a$ and $ B_0^a$ are the
 Lagrange multipliers associated to the following constraints

\begin{eqnarray}
\label{c1}
\varphi^{1a} & = & D_i \pi^{ia} + \epsilon^{abc} {\phi_i}\!^b P^{ic}
= 0  ,\\
\label{c2}
\varphi^{2a} & = & D_i P^{ia} - \epsilon^{abc} {\phi_i}\!^b \pi^{ic}
= 0  ,\\
\label{c3}
\varphi^{3a} & = & \epsilon^{oij} D_{[i} {\phi_{j]}}\!^a = 0  , \\
\label{c4}
\varphi^{4a} & = & \epsilon^{oij} ({F_{ij}}\!^a + \epsilon^{abc}
{\phi_j}\!^b {\phi_i}\!^c) = 0  .
\end{eqnarray}

We may write the algebra of the constraints using a compact notation
in the
 following way
\begin{equation}
\label{algebra}
\{ \varphi^{ia} {\scriptstyle (x)}, \varphi^{jb} {\scriptstyle (x')}
\} = (C_k^{ij})^{abc} \varphi^{kc} {\scriptstyle (x)} \delta^2
{\scriptstyle (x-x')}  .
\end{equation}
where $C_k^{ij}$ are given by
\begin{eqnarray}
(C_k^{1j})^{abc} & = & \delta_k^j \epsilon^{abc}  ,\nonumber  \\
(C_k^{2j})^{abc} & = & m_j \delta_k^{j+m_j} \epsilon^{abc} \; \;
\mbox{ con } \; \; m_j \equiv -(-1)^j  , \nonumber \\
\label{C}
(C_k^{mn})^{abc} & = & 0  ,\nonumber  \\
(C_k^{ij})^{abc} & = & (C_k^{ji})^{abc}  .
\end{eqnarray}
with $ i,j,k = 1,2,3,4 \; $  and  $ \; m,n = 3,4 $.

\noi The eq.(\ref{algebra}) shows that all constraints are first
class and
 independent. We are then  dealing with a closed irreducible
constrained system.

Once the first class constraints  are imposed initially then the
field
equations ensures that they are satisfied at all times.
\noi Moreover from the point of view of the quantum field theory,
when quantizing on a canonical gauge, the configuration space is
restricted to the submanifold defined by the first class constraints.
The interesting property of constraints (\ref{c1}-\ref{c4}) is that (\ref{c3}) and
(\ref{c4}) the only ones restricting $\phi$ and $A$ are related to the
Hitchin \cite{H} equations over Riemann surfaces obtained by
reduction from self dual 4-dim Yang Mills connections. This
equations have a very rich structure related to the Teichm\"uller space of
Riemann surfaces as discussed before and are exactly the field equations obtained by
Bershadsky and Vafa from 2 branes on K3.

It is posible to relate (\ref{c3}-\ref{c4}) to (\ref{hit1}-\ref{hit2}) in a precise way.

The constraint (\ref{c4}) exactly agrees with (\ref{hit1}) when we identify
\begin{eqnarray}
\phi_{z} \equiv \phi_{1} + \dot{\imath} \phi_{2} \\
A_{z} \equiv A_{1} + \dot{\imath} A_{2} \\
\partial_{z} = \partial_{1} - \dot{\imath} \partial_{2}
\end{eqnarray}
where $x^{1}$ and $x^{2}$ are local cartesian coordinates. It follows that (\ref{c3}) may be rewritten as
\begin{equation}
D_{\ov{z}} \phi_{z} - D_{z} \phi_{\ov{z}} = 0 \; .
\end{equation}
It then follows that any solution of Hitchin equations is a solution of (\ref{c3}-\ref{c4}). Conversely, given a solution of (\ref{c3}-\ref{c4}) one may fix the gauge freedom with parameter ${\varepsilon_2}\!^a$ in (\ref{symm}), on a open set over the Riemann surface $\Sigma$, by imposing the gauge fixing condition
\begin{equation}
D_{\ov{z}} \phi_{z} = 0 \; .
\end{equation}
In terms of the local cartesian coordinates, this gauge fixing condition may be rewritten
\begin{equation}
D_1 \phi_1 + D_2 \phi_2 = 0 \; ,
\end{equation}
and can always be imposed on an open set over $\Sigma$. Consequently, over an open set of $\Sigma$ both set of equations are equivalent. The converse argument, however, has not been stablished over the compact Riemann surface $\Sigma$. We will denote $\cal H$ the space of solutions of (\ref{c3}),(\ref{c4}). Having analysed (\ref{c3}) and (\ref{c4}) we may now discuss the complete content of the first class constraints (\ref{c1})-(\ref{c4}). Let $(\phi,A)$ be a point in $\cal H$. We may consider a tangent vector to $\cal H$ at $(\phi,A)$, we will describe it in terms of a variation $(\Delta \phi, \Delta A)$. From (\ref{c3}) and (\ref{c4}) we obtain
\begin{eqnarray}
\epsilon^{oij}[(D_i \Delta A_j)^{a} + \epsilon^{abc} \Delta {\phi_j}\!^b {\phi_i}\!^c] = 0 \\
\epsilon^{oij}[(D_i \Delta \phi_j)^{a} + \epsilon^{abc} \Delta {A_i}\!^b {\phi_j}\!^c] = 0
\end{eqnarray}
If we now identify the dual to the tangent vectors with the canonical momenta to $A$ and $\phi$:
\begin{eqnarray}
\pi^{i} & \equiv & \Delta A_j \epsilon^{oij} \\
P^{i} & \equiv & - \epsilon^{oij} \Delta \phi_j \; ,
\end{eqnarray}
we then obtain precisely the constraints (\ref{c1}) and (\ref{c2}). A solution $(\phi, A, P, \pi)$ of the constraints (\ref{c1})-(\ref{c4}) is then a point in the cotangent bundle $T^{*}\cal H$. To obtain the physical phase space we must still factor out the gauge group. To do so, we notice from (\ref{symm}) that the gauge transformations for $\pi$ and $P$ are the same as the gauge transformations for the variations $\Delta A$ and $\Delta \phi$ respectively, provided the variation of $\varepsilon_1$ is identified to $\varepsilon_4$ and the variation of $\varepsilon_2$ to $\varepsilon_3$. We conclude then that by identifying gauge equivalent elements of $\cal H$, we automatically identify cotangent vectors. The physical phase space is thus $T^{*} (\cal H/G)$ where $\cal G$ denotes the gauge group (\ref{symm}). The Hilbert space of quantum physical states is then the space of square integrable functions on the space $\cal H/G$, which can be identified (up to the point already mentioned) with the space of ``self-dual'' solutions of Hitchin equations, classifying the holomorphic structures of a rank 2 holomorphic vector bundle $V$ of degree zero, with stable pairs $(V, \phi)$. We have thus shown that the physical phase space of the TQFT we proposed can be exactly quantized. For completness we present now the BRST invariant effective action of the theory.

To construct the BRST charge we follow \cite{MGR} and introduce the
minimal sector of the extended phase space expanded by the original
canonical  conjugate pairs $(A_i^a, \pi^{ia}), (\phi_i^a, P^{ia})$,
and by $(C_i^a, \mu^{ia})$ the canonical ghost and anti-ghost
associated to the first class constraints.

The off-shell nilpotent BRST charge is given by

\begin{equation}
\label{ch}
\Omega = \int_{\Sigma} d^2 x \left [ {C_i}\!^a \varphi^{ia}  -  {\scriptstyle
\frac{1}{2}} (C_k^{ij})^{acb} {C_i}\!^b {C_j}\!^c \mu^{ka} \right ] \; ,
\end{equation}
Using the values of the $C^{ij}_k$ given by  (\ref{C}), the BRST
charge can be rewritten as
\begin{eqnarray}
\Omega & = & \int_{\Sigma} d^2 x \left [ {C_i}\!^a \varphi^{ia} + \epsilon^{abc} (
{\scriptstyle \frac{1}{2}} ({C_1}\!^b {C_1}\!^c - {C_2}\!^b
{C_2}\!^c) \mu^{1a} + {C_1}\!^b {C_2}\!^c \mu^{2a} \right .\nonumber \\
  &  &\left . + ({C_1}\!^b {C_3}\!^c - {C_2}\!^b {C_4}\!^c) \mu^{3a} +
({C_2}\!^b {C_3}\!^c + {C_1}\!^b {C_4}\!^c) \mu^{4a} )\right ] \; .
\end{eqnarray}

We may now  introduce the non minimal sector of the extended phase
space \cite{CR}.
It contains extra ghost, antighosts, the C-fields $C_{2i}, C_{3i}$
and the Lagrange multipliers $\lambda_i^0, \theta_i^0$.

\noi The BRST invariant effective action is then given by

\begin{equation}
S_{eff}=\int_{M} d^3x  \left [\pi^i \dot A_i + P^i \dot{\phi_i} + \mu^i \dot
C_i  + \hat{\delta}(\lambda_i^0 \mu^i) +  \hat{\delta}(C_{2i}
\chi_2^i)\right ] \; .
\end{equation}
where $\chi_2^i$ are the gauge fixing conditions associated to the first class constraints $\varphi^{i}$, and $\lambda_i^0$ its corresponding Lagrange multipliers.

The BRST transformation for the canonical variables is given by
\begin{equation}
\hat{\delta}Z={(-1)^{\epsilon_Z}} \{ Z,\Omega \},
\end{equation}
where $\epsilon_Z$ is the grassmanian parity of $Z$. Thus we obtain
for the original fields the following BRST transformation rules:

\begin{eqnarray}
\hat{\delta} {A_\mu}\!^a & = & - D_\mu {C_1}\!^a  +  \epsilon^{abc}
{\phi_\mu}\!^b \, {C_2}\!^c \; ,\nonumber  \\
\hat{\delta} {\phi_\mu}\!^a & = & - D_\mu {C_2}\!^a - \epsilon^{abc}
{\phi_\mu}\!^b \, {C_1}\!^c \; , \nonumber \\
\hat{\delta} {\eta_\mu}\!^a & = & - D_\mu {C_3}\!^a + \epsilon^{abc}
{\phi_\mu}\!^b \, {C_4}\!^c - \epsilon^{abc} {\eta_\mu}\!^b \,
{C_1}\!^c + \epsilon^{abc} {B_\mu}\!^b \, {C_2}\!^c \; ,\nonumber  \\
\hat{\delta} {B_\mu}\!^a    & = & - D_\mu {C_4}\!^a - \epsilon^{abc}
{\phi_\mu}\!^b \, {C_3}\!^c - \epsilon^{abc} {\eta_\mu}\!^b \,
{C_2}\!^c - \epsilon^{abc} {B_\mu}\!^b \, {C_1}\!^c \; .
\end{eqnarray}
While for the ghosts and anti-ghosts the BRST transformation rules are:
\begin{eqnarray}
\hat{\delta} {C_1}\!^a & = & - {\scriptstyle \frac{1}{2}}
\epsilon^{abc}({C_1}\!^b {C_1}\!^c - {C_2}\!^b {C_2}\!^c) \;
,\nonumber  \\
\hat{\delta} {C_2}\!^a & = & - \epsilon^{abc} {C_1}\!^b {C_2}\!^c \;
,\nonumber  \\
\hat{\delta} {C_3}\!^a & = & - \epsilon^{abc} ({C_1}\!^b {C_3}\!^c -
{C_2}\!^b {C_4}\!^c) \; , \nonumber \\
\hat{\delta} {C_4}\!^a & = & - \epsilon^{abc} ({C_2}\!^b {C_3}\!^c +
{C_1}\!^b {C_4}\!^c ) \; ,\nonumber \\
\hat{\delta} \mu^{1a} & = & - \varphi^{1a} - \epsilon^{abc} {C_i}\!^b
\mu^{ic} \; , \nonumber \\
\hat{\delta} \mu^{2a} & = & - \varphi^{2a} - \epsilon^{abc}
 ({C_1}\!^b \mu^{2c} - {C_2}^b \mu^{1c} + {C_3}\!^b \mu^{4c} -
{C_4}\!^b \mu^{3c}) \; ,\nonumber \\
\hat{\delta} \mu^{3a} & = & - \varphi^{3a} - \epsilon^{abc}
({C_1}\!^b \mu^{3c} + {C_2}\!^b \mu^{4c} ) \; ,\nonumber  \\
\hat{\delta} \mu^{4a} & = & - \varphi^{4a} - \epsilon^{abc}
({C_1}\!^b \mu^{4c} - {C_2}\!^b \mu^{3c} ) \; .
\end{eqnarray}
The Hilbert space of physical states is then defined as the cohomology classes of BRST charge (\ref{ch}) acting on the space of functionals on the phase space. They must correspond then to the space of functionals on $T^{*}(\cal H/G)$. It is interesting that such mathematical result arises directly from the equivalence of both quantization procedures. In the first one we solved completely the constraints and then quantize on the resulting physical space, in the latest one we imposed the constraints on the functionals of the original phase space, giving rise to the cohomology classes of the BRST charge.

\section{Monopole Solutions over $S^2$ to the field equations}
\setcounter{equation}{0}

An interesting property of the field equations (\ref{feq}-\ref{feq2})  is that the
connection $A$ describing the Dirac's monopole  over the sphere $S^2$, that is
the Hopf fibration, is a static solution of them with the time component of $A$ and $\phi$ being zero.

We only need to  look  for solutions of the field equations
(\ref{feq1}) and (\ref{feq2}) over $S^2$, that we will rewrite below without the
internal indices
\begin{eqnarray}
\label{1}
F_{\mu \nu} + [\phi_\nu,\phi_\mu] = 0\; ,\\
\label{2}
D_{[\mu}\phi_{\nu]} = 0 \;.
\end{eqnarray}

We consider an  ansatz for the solution, which may  be written in
terms of differential forms as:

\begin{eqnarray}
A^1 & = & A^2 = 0 \; ,\nonumber \\
A^3 & = & n (1-cos \theta) d\phi \; ,
\end{eqnarray}
and
\begin{eqnarray}
\phi^1 & = & f(\theta) cos \phi d\theta - g(\theta) sen \phi d\phi \;
, \nonumber \\
\phi^2 & = &  f(\theta) sen \phi d\theta + g(\theta) cos \phi d\phi
\; , \nonumber\\
\phi^3 & = & 0 \; .
\end{eqnarray}
Where the  functions $f(\theta)$ and $g(\theta)$ have to be
determined from the field equations.
After solving all the field equations, we obtain for them:

\begin{eqnarray}
g'(\theta)& = &(1+n-n cos \theta)f(\theta) \; ,\nonumber \\
f(\theta) g(\theta) & = & n sen \theta \; ,
\end{eqnarray}
where the prime denotes derivative respect to $\theta$.
Finally we obtain the following solution for the square of $g(\theta)$
\begin{equation}
g^2(\theta)=[\frac{1}{2}n^2 cos 2\theta - 2n(n+1)cos \theta]+C ,
\end{equation}
where $C$ is an integration constant, which must ensure the
positiveness of $g^2$.

The curvature 2-form $F$ satisfies
\begin{equation}
\int_{S^2} F^3 = 4 \pi n ,
\end{equation}
where $F^3$ refers to the $SU(2)$ 3-index component of $F$. The solution so
obtained for $F$ characterizes the connection 1-form of the Dirac's magnetic
monopole.
This solution shows a definite distinction between the topological theory we are considering and the BF TFT, caracterized by flat connections only.
\section{Conclusions}

We introduced an extension of the BF topological field action. The
theory describes the evolution of the space of solutions of Hitchin
self-dual equations over topologically non-trivial Riemann surfaces.
The functional integral of the generalized BF action defines then a
canonical map between the holomorphic structures of rank 2 vector
bundles of degree zero over a Riemann surface of genus $g >1$. The
theory allows then a realization for the particular problem
considered, of the general approach proposed by Atiyah \cite{A}.

The physical phase space of the theory was exactly quantized in terms of the cotangent bundle $T^{*}(\cal H/G)$. The Hilbert space of quantum physical states may consequently be identified to the space of square integrable functionals on the space $\cal H/G$ described in section 3. The nilpotent BRST charge was also constructed and it was concluded that the above mentioned Hilbert space must be equivalent to the space of cohomology classes of the BRST operator. Since one sector of the space $\cal H/G$ provides a model of Teichm\"{u}ller space which is isomorphic to the space ${\cal A/G}_{1}$, where $\cal A$ are the flat $SO(2,1)$ connections and ${\cal G}_{1}$ the gauge group arising in topological gravity, we then conclude that the Hilbert space of physical states describing topological gravity \cite{Ge} is a subspace of the corresponding one describing self-dual field configurations arising from the dimensional reduction of 4-dim self dual Yang-Mills.

Finally we found, over $S^2$, a static class of solutions to the field equations of the TFT. They describe Dirac monopoles over $S^2$ showing an explicit distinction with respect to the flat connections arising from the BF TFT.

The TFT we have presented may be generalized in a straightforward way to higher dimensions, in particular to 5 dim. However it is not clear that such generalization will describe a TFT with physical phase space related to 4-dim self-dual Yang-Mills fields (as the case we have presented) or even better to the 4-dim $U(1)$ monopole equations introduced by Witten \cite{MGR} following the Seiberg-Witten \cite{SW} duality approach. This problem is under study.

\end{document}